\documentclass{mn2e}
\usepackage{epsf}
\usepackage{times}
\newcommand{\etal}{{et al}\/.}
\begin{document}
\title[The giant lobes of Centaurus A]{High-energy particle acceleration and production of ultra-high-energy
cosmic rays in the giant lobes of Centaurus A}
\author[M.J.~Hardcastle \etal]{M.J.\ Hardcastle$^1$, C.C. Cheung$^2$,
I.J. Feain$^3$ and \L. Stawarz$^{4,5}$\\
$^1$ School of Physics,
  Astronomy and Mathematics, University of
Hertfordshire, College Lane, Hatfield, Hertfordshire AL10 9AB, UK\\
$^2$ NASA Goddard Space Flight Center, Astrophysics Science Division,
  Code 661, Greenbelt, MD, 20771, USA\\
$^3$ CSIRO Australia Telescope National Facility, P.O. Box 76, Epping,
  NSW 1710, Australia\\
$^4$ Kavli Institute for Particle Astrophysics and Cosmology, Stanford
  University, Stanford, CA 94305, USA\\
$^5$ Obserwatorium Astronomiczne, Uniwersytet Jagiello\'nski, ul. Or{\l}a
171, PL-30244 Krak\'ow, Poland
}
\maketitle
\begin{abstract}
The nearby radio galaxy Centaurus A is poorly studied at high
frequencies with conventional radio telescopes because of its very
large angular size, but is one of a very few extragalactic objects to
be detected and resolved by the {\it Wilkinson Microwave Anisotropy
Probe} ({\it WMAP}). We have used the five-year {\it WMAP} data for
Cen~A to constrain the high-frequency radio spectra of the 10-degree
giant lobes and to search for spectral changes as a function of
position along the lobes. We show that the high-frequency radio
spectra of the northern and southern giant lobes are significantly
different: the spectrum of the southern lobe steepens monotonically
(and is steeper further from the active nucleus) whereas the spectrum
of the northern lobe remains consistent with a power law. The inferred
differences in the northern and southern giant lobes may be the result
of real differences in their high-energy particle acceleration
histories, perhaps due to the influence of the northern middle lobe,
an intermediate-scale feature which has no detectable southern
counterpart. In light of these results, we discuss the prospects for
{\it Fermi Gamma-ray Space Telescope} detections of inverse-Compton
emission from the giant lobes and the lobes' possible role in the
production of the ultra-high energy cosmic rays (UHECR) detected by
the Pierre Auger Observatory. We show that the possibility of a {\it
Fermi} detection depends sensitively on the physical conditions in the
giant lobes, with the northern lobe more likely to be detected, and
that any emission observed by {\it Fermi} is likely to be dominated
by photons at the soft end of the {\it Fermi} energy band. On the other hand we
argue that the estimated conditions in the giant lobes imply that
UHECRs can be accelerated there, with a potentially detectable
$\gamma$-ray signature at TeV energies.
\end{abstract}

\section{Introduction}
\label{intro}

Centaurus A is the closest radio galaxy to us (we adopt $D = 3.7$ Mpc,
the average of 5 distance indicators in Ferrarese \etal\ 2007). Its
proximity makes it one of the brightest extragalactic radio sources in
the sky at low frequencies (only exceeded by Cygnus A: Baars \etal\
1977), but also means that the outer `giant' double lobes (throughout
the paper we use the nomenclature adopted by Alvarez et al 2000)
subtend an angle of $\sim 10^\circ$ on the sky, although their total
physical size ($\sim 600$ kpc in projection) is not unusually large
for an radio galaxy of Cen A's luminosity. The large angular size of
the lobes has prevented the type of {\it spatially resolved},
multifrequency study of their spectral structure that is commonplace
for more distant radio galaxies (e.g. Alexander \& Leahy 1987). While
detailed low-frequency maps of the giant lobes have been available for
many years (e.g. Cooper, Price \& Cole 1965), published radio data from
ground-based observations only exist up to 5 GHz (Junkes \etal\ 1993)
and the spectral study of Alvarez \etal\ (2000), involving (at many
frequencies) painstaking graphical integration of contour maps, was
only able to determine overall spectra for the giant lobes, finding no
evidence for deviation from a single power law between 408 MHz and 5
GHz. The low-frequency two-point spectral index maps of Combi \&
Romero (1997), however, do show some evidence for position-dependent
spectral steepening, particularly towards the end of the southern
giant lobe.

Cen~A is widely believed to be a
restarting radio galaxy, in the sense that the inner lobes are the
result of the current nuclear activity, while the giant outer lobes
are the result of a previous outburst (Morganti \etal\ 1999). This
picture is supported by the observation of hot thermal X-ray emission,
apparently the result of strong shocks, surrounding both inner lobes
(Kraft \etal\ 2003, 2007; Croston \etal , in prep.) which implies that
they are propagating supersonically into the intergalactic medium
(IGM) of Cen~A and are disconnected from the giant lobes. However, the
nature of the intermediate-scale northern middle lobe (NML: Morganti
\etal\ 1999) is not completely clear in this model. In standard
spectral ageing models (e.g. Jaffe \& Perola 1973), we
might expect to see spectral steepening at high frequencies in the
giant lobes; a measurement of spectral ageing gives a model-dependent
constraint on the time since the last injection of high-energy
electrons into these lobes. Such a constraint on spectral age could be
compared with other estimates of the dynamical age of the radio
source, and would therefore be of considerable interest, but the work
of Alvarez \etal\ (2000) only sets upper limits on this quantity,
since they do not see any deviation from a power-law spectrum.

\begin{figure*}
\epsfxsize 17.5cm
\epsfbox{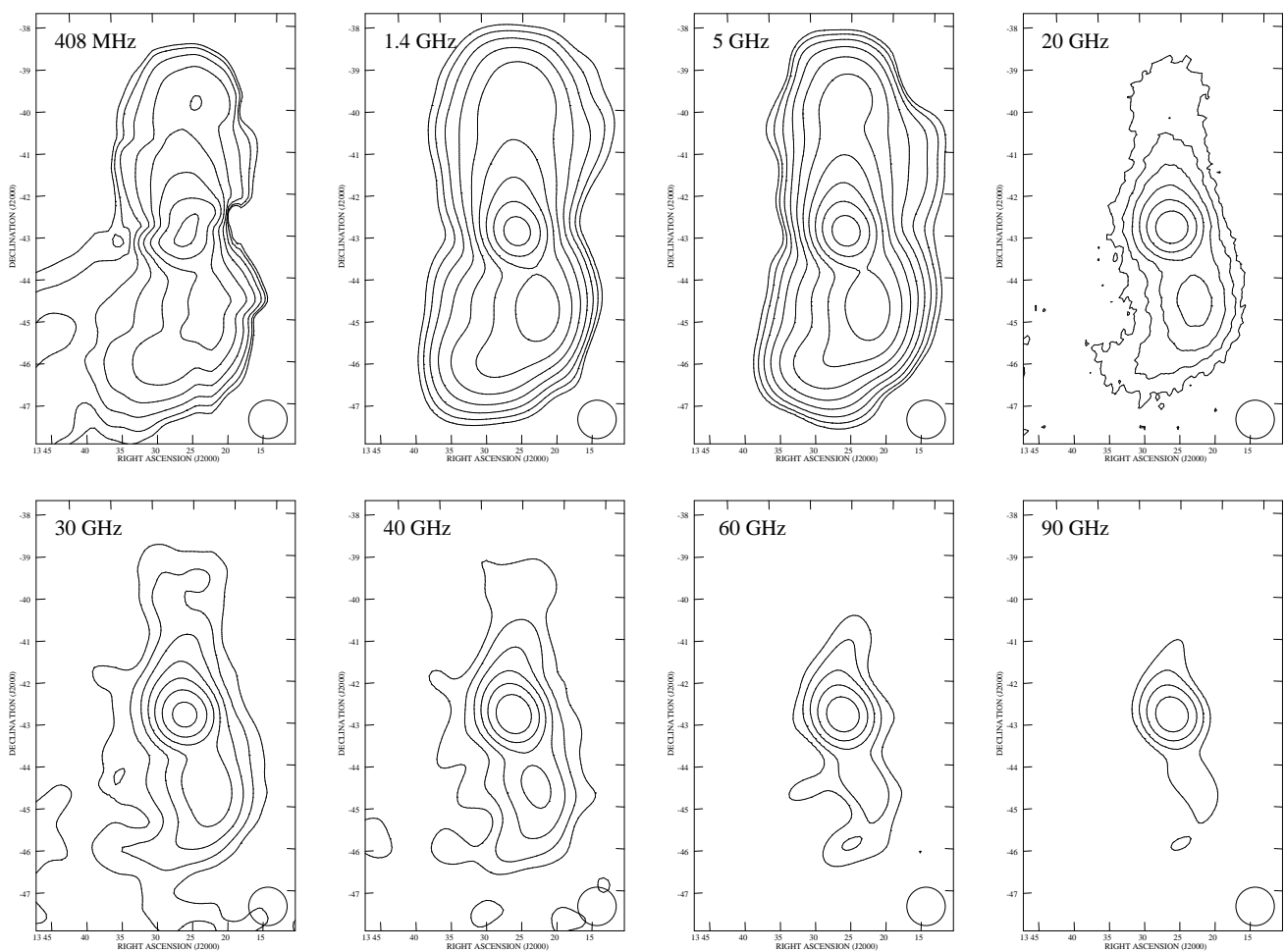}
\caption{Large-scale structure of Centaurus A with a resolution of
  $\sim 0.83^\circ$. Contours are at $1,2,4\dots$ times the base level
  specified for each map, and take no account of background, except
  for the 408-MHz map, from which a constant background of 43 Jy
  beam$^{-1}$ has been subtracted. Top row, from left to right: 408
  MHz (3 Jy beam$^{-1}$), 1.4 GHz (1 Jy beam$^{-1}$), 5 GHz (0.25 Jy
  beam$^{-1}$), 20 GHz (1.0 Jy beam$^{-1}$). Bottom row, from left to
  right: 30 GHz (0.5 Jy beam$^{-1}$), 40 GHz (1.0 Jy beam$^{-1}$), 60
  GHz (1.0 Jy beam$^{-1}$), 90 GHz (2.0 Jy beam$^{-1}$). Circles in
  the bottom right-hand corner of each image indicate the beam size
  (diameter shows FWHM). The 20-GHz data have different noise
  characteristics from the other {\it WMAP} images because they have
  not been convolved with a Gaussian; instrumental noise on scales
  smaller than the effective beam is therefore visible. See the text
  (Section \ref{wmap-pro}) for discussion of the convolution,
  effective resolution and beam area of these images.}
\label{convolved-data}
\end{figure*}

The giant lobes of Cen~A are also interesting because they are
predicted to be strong sources of inverse-Compton emission as the
relativistic electrons in the lobes scatter cosmic microwave
background (CMB) photons to high energies; a detection of
inverse-Compton emission from Cen~A would constrain the magnetic field
strength in the lobes of Fanaroff \& Riley (1974) class I (hereafter
FRI) radio sources in general: we have little information on the
magnetic field strengths in these low-power radio galaxies at present.
However, X-ray emission from this process would be distributed on
similar scales to the giant lobes, making it hard to detect. Cooke,
Lawrence \& Perola (1978) claimed an early detection of the giant
lobes using {\it Ariel V}, but Marshall \& Clark (1981) argued that
this was the result of point source contamination, placing a much
lower upper limit on the flux from {\it SAS 3} observations. At soft
X-ray energies (e.g. Arp 1994) the situation is seriously confused by
the presence of known thermal X-ray emission from the interstellar
medium of the host galaxy, which more recently has been extensively
studied with {\it Chandra} and {\it XMM-Newton} (e.g.\ Kraft \etal\
2003), and is also hard to study because the X-ray emission fills the
field of view of modern soft X-ray imaging instruments such as {\it
ROSAT} (Arp 1994), {\it XMM}, and {\it ASCA} (Isobe \etal\ 2001),
presenting almost insuperable problems of background modelling and
subtraction. However, the spectrum of the inverse-Compton emission
should be hard up to high energies (exactly how high depends on the
model adopted for the electron energy spectrum, as we will discuss
below). There are thus also interesting constraints from observations
at MeV to GeV energies made with the {\it Compton Gamma-Ray
Observatory} (e.g. Steinle \etal\ 1998; Sreekumar \etal\ 1999), which
do not detect the giant lobes but again set upper limits on their
high-energy flux densities. More sensitive hard X-ray/$\gamma$-ray
observations exist with wide-field instruments like {\it INTEGRAL}
(e.g. Rothschild \etal\ 2006) and the {\it Swift} Burst Alert
Telescope (e.g. Markwardt \etal\ 2005) but as these are coded-aperture
instruments they have limited sensitivity to extended emission (see
e.g. Renaud \etal\ 2007). At present, therefore, there is no
unambiguous detection of X-ray or $\gamma$-ray inverse-Compton
emission from the giant lobes of Cen~A. One of us has shown (Cheung
2007) that {\it Fermi}\footnote{Formerly known as the {\it Gamma-ray
Large Area Space Telescope}, {\it GLAST}.} may have the sensitivity to
detect inverse-Compton emission off the CMB from the lobes of Cen~A at
energies from $\sim$100 MeV to 10 GeV. However, the details of this
depend on modelling of the electron energy spectrum at high energies,
which in turn depends on high-frequency radio data.

Finally, Cen~A's giant lobes are possible sources of ultra-high energy
cosmic rays (UHECRs). Cen~A's proximity means that all aspects of the
active galaxy -- central AGN, inner jets and lobes, and giant lobes --
have long been considered as possible UHECR accelerators (see e.g.
Cavallo 1978; Romero \etal\ 1996, and, more recently, Gureev \& Troitsky 2008 and references therein). Interest in
Cen~A has been spurred by the remarkable discovery that 2 of the 27
UHECR events detected so far by the Pierre Auger Observatory
(hereafter `PAO'; Abraham \etal\ 2007) appear to be arriving from the
direction of the centre of Cen~A, while at least 2 additional events
may be associated with it (e.g. Gorbunov \etal\ 2008a; Wibig \& Wolfendale
2007; Fargion 2008) due to the large angular extent of the giant radio
lobes (Gorbunov \etal\ 2008b; Moskalenko \etal\ 2008). Most scenarios
discussed in the literature to date assume that UHECRs are produced
near the supermassive black hole (SMBH) or in the inner jets (e.g.
Cuoco \& Hannestad 2008; Kachelriess, Ostapchenko \& Tomas 2008), but
an explanation in terms of the giant lobes has the advantage that it
can easily account for the PAO events seen on larger scales. In order
to investigate this quantitatively we need information about the
magnetic field strengths and the {\it leptonic} particle acceleration
in these lobes, which can be provided by a combination of
high-frequency radio observations and inverse-Compton constraints or
measurements.

The {\it Wilkinson Microwave Anisotropy Probe} ({\it WMAP}) has
observed the whole sky at frequencies around 20, 30, 40, 60 and 90 GHz
(known as K, Ka, Q, V and W bands respectively) with the aim of
measuring structure in the CMB (e.g.\ Hinshaw \etal\ 2008). The
currently available {\it WMAP} data represent 5 years of observations.
Cen~A is clearly detected, and spatially resolved, in the {\it WMAP}
observations at all frequencies (e.g. Page \etal\ 2007) and Israel
\etal\ (2008) have recently presented {\it WMAP}-derived measurements
of the flux density of the whole source, showing that there is clear
steepening in the integrated spectrum at high frequencies. Thus the
data are available to carry out a study of the variation of the radio
spectrum as a function of position, to fit spectral ageing models to
the large-scale lobes and investigate whether we can learn anything
about the source dynamics, and to make predictions of the expected
inverse-Compton emission from the giant lobes.

In this paper we present the results of such a study. We first combine
the 5-year {\it WMAP} data on Cen~A with single-dish radio images at
lower frequencies to make spatially resolved measurements of the radio
spectra from 408 MHz to 90 GHz. We then discuss the implications of
the high-frequency detections for the dynamics of Cen~A, for possible
inverse-Compton detections of the giant lobes at high energies, and
for acceleration of UHECRs and their possible $\gamma$-ray emission
signatures.

\section{Data}

\subsection{Radio data}

We obtained electronic versions of ground-based radio maps
of Cen~A to provide low-frequency counterparts to the five {\it WMAP}
images. At 408 MHz we used the all-sky map of Haslam \etal\
(1982) which is available from the NCSA Astronomy Digital Image
Library\footnote{See http://adil.ncsa.uiuc.edu/document/95.CH.01.01 .}.
The 1.4-GHz and 4.75-GHz (hereafter, 5-GHz) data were taken with the Parkes telescope and kindly
provided to us by Norbert Junkes; the 5-GHz data are those presented
by Junkes \etal\ (1993).

\subsection{WMAP data and processing}
\label{wmap-pro}

FITS maps of the 5-year WMAP data for Cen~A at all 5 {\it WMAP}
frequencies were kindly provided by N. Odegard. For each map, a
monopole defined outside of the {\it WMAP} Kp2 mask (see e.g.\ Hinshaw \etal\
2007) and the internal linear combination (ILC) CMB map were
subtracted. The data were interpolated from HEALPix to a Galactic
co-ordinate grid with $0.1^\circ$ spacing. The images were then
converted into a celestial co-ordinate frame using AIPS, and the units
of the maps converted from mK to Jy beam$^{-1}$ for ease of comparison
with the ground-based radio maps, which have these units. For the unit
conversion we used the beam areas tabulated by Page \etal\ (2003) and
added a `beam size' header to the FITS file that defined the beam to
be a Gaussian of the same area. Although the WMAP beam is not a
Gaussian (Page \etal), this approximation has no effect on our flux
density measurements, since the two identical values of beam area
cancel; in addition, the Gaussian FWHM of the beam calculated in this
way is similar to the true FWHM tabulated by Page \etal\ (2003), and
so gives us a convenient way to characterize the resolution of the
resulting maps. We were then able to use AIPS for simple analysis of
the WMAP data and software based on the Funtools
package\footnote{Documented at
http://hea-www.harvard.edu/saord/funtools/ ; see also Mandel, Murray
\& Roll (2001).} for measurement of flux densities from matched
regions of the ground-based and {\it WMAP} radio maps. The effective
central frequencies of the {\it WMAP} bands for extended emission with
a spectral index of 0.7 are 22.5, 32.7, 40.4, 60.1 and 92.9 GHz (based
on the bandpass characterization of Jarosik \etal\ 2003) and we treat
measured flux densities as being measured at these central frequencies
in all subsequent analysis, although for simplicity we will refer to
the different bands as the 20, 30, 40, 60 and 90-GHz bands in what
follows.

In measuring flux densities from regions of Cen~A with {\it WMAP} we
need to take careful account of the foreground and background
characteristics of the data. Firstly, the ILC subtraction described
above only takes account of structure in the CMB on scales greater
than $1^\circ$: this gives rise to real structure on small scales in
the maps of this $10^\circ$ source, which is a problem at higher
frequencies and correspondingly high resolutions. Secondly,
synchrotron emission from the Milky Way provides a foreground which is
both position and frequency-dependent, being strongest at low
frequencies and towards the south of the Cen~A field (e.g. de
Oliviera-Costa \etal\ 2008).

For our analysis we convolved all the images with a Gaussian that gave
us approximately the resolution of the 20-GHz WMAP image
(0.83$^\circ$) at all observing frequencies, with the exception of the
20-GHz data themselves and of the 408-MHz dataset, which already has a
very similar intrinsic resolution. The convolution has two effects on
our analysis: for the {\it WMAP} data, it helps to reduce any
contamination from the unsubtracted real structure in the CMB; and, for
all the datasets, it allows us to measure flux densities from extended
regions with sufficient confidence that we are seeing the same structures at
different frequencies.

Fig.\ \ref{convolved-data} shows a montage of the radio images used in
this paper. Full-resolution {\it WMAP} images are presented by Israel
\etal\ (2008).

\section{Results}

\subsection{Spectral index mapping}

\begin{figure}
\epsfxsize 8.5cm
\epsfbox{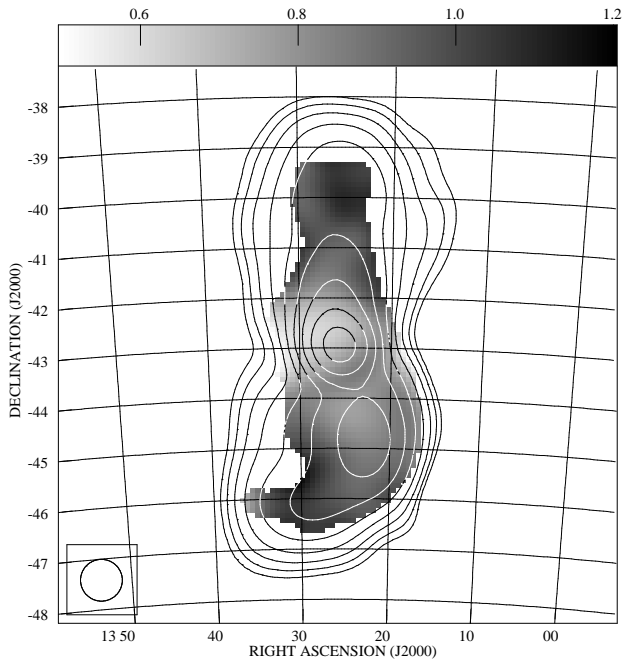}
\caption{Map of two-point spectral index $\alpha$ between 1.4 and 30 GHz in the
  giant lobes of Cen~A. Only pixels where the total intensity exceeds
  5 times the off-source r.m.s. level are plotted. Contours are from the 1.4-GHz Parkes
  data at $1, 2, 4\dots$ Jy beam$^{-1}$. The resolution (0.83$^\circ$)
  is indicated by the circle in the bottom left-hand corner.}
\label{spix}
\end{figure}

To search for evidence for spectral differences in the large-scale
structure of Cen~A as a function of position we initially constructed
a map of the two-point spectral index ($\alpha =
-\log(S_1/S_2)/\log(\nu_1/\nu_2)$) between 1.4 and 30 GHz (Fig.
\ref{spix}). Such a map can only be approximate given that the beams
of the ground-based and {\it WMAP} data are only approximately
matched, and it cannot take account of the varying position-dependent
background in the {\it WMAP} images, but it does provide an indication
of whether the spectral steepening already observed by Israel \etal\
(2008) is position-dependent. The spectral index map gives a strong
indication that the spectral index does indeed steepen as a function
of distance from the `nucleus' (which at this resolution includes the
entire inner lobe and north middle lobe structures as well as the
flat-spectrum core), confirming the lower-frequency (0.4--1.4 GHz)
results of Combi \& Romero (1997).

\begin{figure}
\epsfxsize 8.5cm
\epsfbox{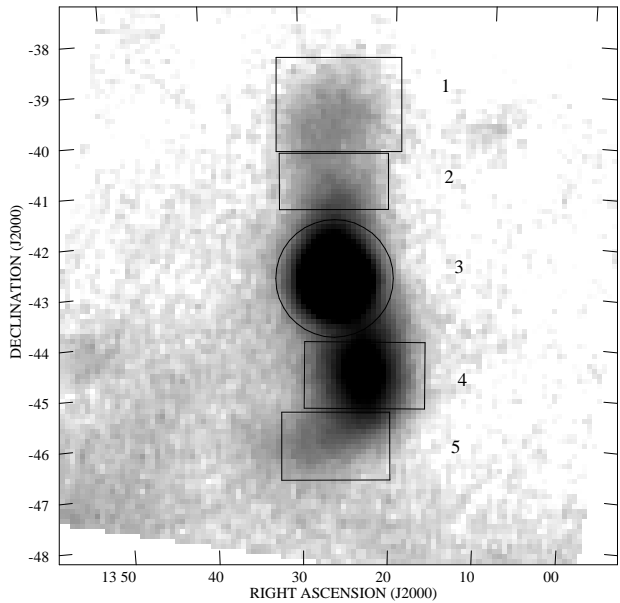}
\caption{Spectral extraction regions used for
 Table \ref{fluxes} and Fig.\ \ref{spectrum}. The
  greyscale shows the 20-GHz {\it WMAP} data.}
\label{regions}
\end{figure}

\subsection{Flux density measurements}

\begin{figure}
\epsfxsize 8.5cm
\epsfbox{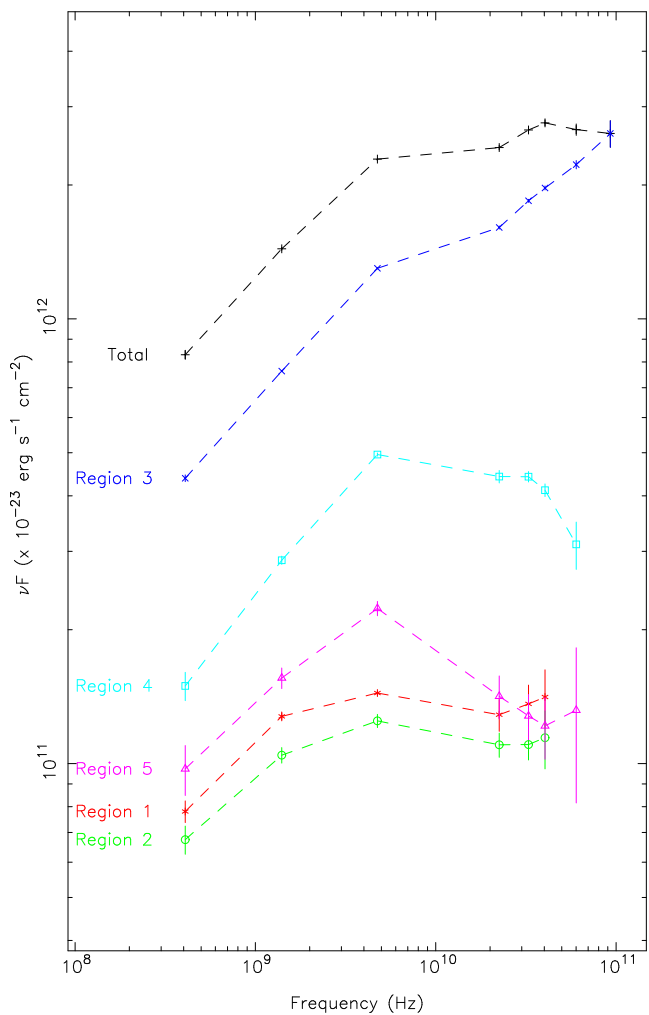}
\caption{Flux density $\times$ frequency ($\nu f_\nu$) for the regions
  of Table \ref{fluxes} (see Fig.\ \ref{regions}) as a
  function of frequency. 1 Jy Hz is $10^{-23}$ erg s$^{-1}$ cm$^{-2}$.}
\label{spectrum}
\end{figure}

To investigate spectral changes in the giant lobes more quantitatively
we next divided the radio source into 5 regions encompassing all of
the structure detected at 20 GHz. These regions are shown in Fig.\
\ref{regions} and for simplicity we number them 1--5 from north to
south on the source. Regions 1 and 2 are respectively the outer and
inner regions of the northern giant lobe, region 3 is dominated by the
emission from the nucleus, inner lobes and north middle lobe, and
regions 4 and 5 are respectively the inner and outer regions of the
southern giant lobe. The regions are all substantially larger than the
adopted {\it WMAP} beam sizes. For each region we defined two
background regions lying adjacent to the source regions in the
east-west direction. The background regions were used both to find a
mean background flux density, which was
subtracted from the total in the source regions (thus removing
contamination from our own Galaxy, whose strength varies with position
on the maps) and to determine an off-source r.m.s. for each map, from
which we estimated the on-source noise. Flux measurements and
background subtraction were carried out in the same way for all the
ground-based and {\it WMAP} datasets. The measurements of flux
density for each region and frequency, together with the error
estimated from the background region r.m.s., are tabulated in Table
\ref{fluxes} and plotted together in Fig.\ \ref{spectrum}.

\begin{table*}
\caption{Flux densities of the regions of Cen~A as a function of
  frequency. The final column gives the sum of all measured quantities
  in the preceding columns. Blanks indicate combinations of frequency
  and region where significant emission was not visible by eye.}
\label{fluxes}
\begin{tabular}{rrrrrrr}
\hline
Frequency&\multicolumn{6}{c}{Flux density (Jy)}\\
(GHz)&Region 1&Region 2&Region 3&Region 4&Region 5&Total\\

\hline
0.408 & $191.3 \pm 10.8$ & $165.3 \pm 12.0$ & $1074.0 \pm 18.8$ & $366.3 \pm 26.7$ & $238.5 \pm 30.7$ & $2035.2 \pm 47.6$ \\
1.4 & $91.2 \pm 2.1$ & $74.6 \pm 3.0$ & $545.0 \pm 1.3$ & $204.7 \pm 4.3$ & $111.2 \pm 6.0$ & $1026.7 \pm 8.3$ \\
4.75 & $30.3 \pm 0.3$ & $26.3 \pm 0.9$ & $273.5 \pm 1.0$ & $104.3 \pm 0.9$ & $47.0 \pm 1.7$ & $481.4 \pm 2.4$ \\
22.5 & $5.7 \pm 0.5$ & $4.9 \pm 0.3$ & $71.3 \pm 0.6$ & $19.6 \pm 0.6$ & $6.3 \pm 0.7$ & $107.9 \pm 1.2$ \\
32.7 & $4.2 \pm 0.4$ & $3.4 \pm 0.3$ & $56.4 \pm 0.4$ & $13.5 \pm 0.4$ & $3.9 \pm 0.5$ & $81.3 \pm 0.9$ \\
40.4 & $3.5 \pm 0.5$ & $2.8 \pm 0.4$ & $48.8 \pm 0.5$ & $10.2 \pm 0.3$ & $3.0 \pm 0.5$ & $68.3 \pm 1.0$ \\
60.1 & -- & -- & $37.0 \pm 0.8$ & $5.2 \pm 0.6$ & $2.2 \pm 0.8$ & $44.3 \pm 1.3$ \\
92.9 & -- & -- & $28.1 \pm 1.9$ & -- & -- & $28.1 \pm 1.9$ \\

\hline
\end{tabular}
\end{table*}

We initially note that the total flux density listed in Table
\ref{fluxes}, the sum of all measured flux densities from the regions,
agrees well with the values tabulated by Israel \etal\ (2008) (their
table 1) within the uncertainties. Our flux densities are generally
slightly lower, and the discrepancy is greatest at the higher
frequencies where we have not attempted to measure fluxes for the
whole source, but even here the differences do not exceed the errors
quoted by Israel \etal\ (larger than ours because they do not restrict
themselves to regions where emission is clearly detected). The
agreement between the two independent analyses strengthens the case
that the flux density measurements are correct. We confirm the
spectral steepening of the integrated source spectrum detected by
Israel \etal

Examination of the SEDs for individual regions (Fig.\ \ref{spectrum})
shows clear differences between the northern giant lobe (regions 1 and
2), the central region (3), and the southern giant lobe (4 and 5). The
central region shows no evidence for spectral steepening except
between 5 and 20 GHz. At {\it WMAP} frequencies the spectral index
$\alpha$ is very consistently about 0.65, close to the lower-frequency
spectral index of 0.70 derived for the inner lobes by Alvarez \etal\
(2000). Given that the central region (region 3) in our analysis
includes the flat-spectrum core (which is strongly variable, e.g.
Israel \etal\ 2008), the inner lobes, the north middle lobe and some
extended lobe emission, it is not surprising that its spectrum is
poorly fitted by a single power-law or curved spectrum. Probably it
consists of (at least) a flat-spectrum component and a component with
a spectral turnover between 5 and 20 GHz.

The spectra of the southern giant lobe regions (regions 4 and 5)
remain relatively flat up to 5 GHz, but show clear steepening at high
frequencies. Moreover, the spectrum of region 5 (the southern half of
the southern giant lobe) is systematically
steeper than that of region 4 (the northern half), even at low frequencies (as indicated
by our spectral index mapping and also seen by Combi \& Romero 1997).
This difference is statistically significant: considering just the
spectral indices between 1.4 and 5 GHz, $\alpha_4 = 0.55 \pm 0.02$
while $\alpha_5 = 0.71 \pm 0.05$. The systematic increase in spectral
curvature as a function of frequency, seen particularly in region 4
(northern south giant lobe),
is characteristic of spectral ageing models, and we explore fits of
such models to the data below.

On the other hand, the northern giant lobe regions (regions 1 and 2)
show no significant spectral differences. Their spectra both start to
steepen between 1.4 and 5 GHz and then remain roughly constant (at
around $\alpha = 1.0$) at higher frequencies. These spectra are
clearly different from those of the southern giant lobe. It does not
seem possible to explain this difference in terms of inadequate
background subtraction; the Galactic background that we subtract off
is a power law, so that it is hard to see how any combination of
partially subtracted background and source could produce the
differences in northern and southern regions. The statistical
differences (given our adopted errors) between the northern and
southern giant lobes are highly significant both at low and high
frequencies. We explore the reasons for this difference below (Section
\ref{giant}).

\begin{figure}
\epsfxsize 8.5cm
\epsfbox{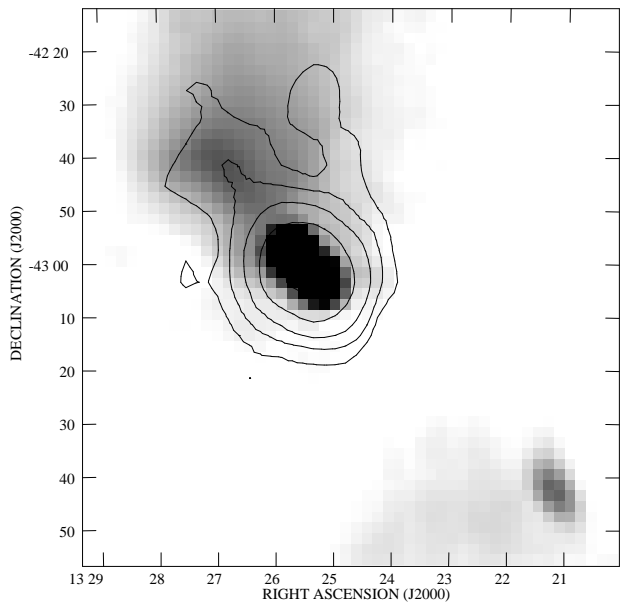}
\caption{{\it WMAP} and radio images of the central parts of Cen~A,
  showing the elongated structure that we associate with the northern
  middle lobe. Contours are from the full-resolution {\it WMAP} 90-GHz
  image, at $0.75 \times (1, 2, 4\dots)$ Jy beam$^{-1}$) overlaid on
  the 5-GHz radio map of Junkes \etal\ (1993). The area of the {\it
  WMAP} beam at this frequency corresponds to an effective resolution
  around 15 arcmin. The compact feature to the SW of the inner
  lobes/core of Cen~A, undetected at 90 GHz, is a background radio
  galaxy.}
\label{nml-fig}
\end{figure}

\subsection{High-resolution imaging of the NML}

Finally we note that in the 90-GHz {\it WMAP} image, the only one to
have the required resolution to distinguish between the inner lobes
and the NML, there is a clear extension of the central point source
(Fig.\ \ref{nml-fig}) which is exactly coincident with the NML. Its
flux density at 90 GHz is around 1.5 Jy which implies a comparatively
flat spectrum ($\alpha \approx 0.6$) between 5 and 90 GHz, measuring
flux from matched regions of the full-resolution 5-GHz map of Junkes
\etal\ (1993) and the 90-GHz {\it WMAP} image. This strongly suggests
that particle acceleration is either ongoing there or has recently
ceased, although given that we only have two spectral data points we
cannot rule out the possibility that the spectrum is flatter simply
due to a higher magnetic field strength (cf. Katz-Stone, Rudnick \&
Anderson 1993). We discuss the implications for the nature of the NML
in Section \ref{nml}.

\section{The nature of the NML}
\label{nml}

As discussed in Section \ref{intro}, the northern middle lobe (NML:
e.g. Morganti \etal\ 1999) is a feature at the base of the northern
giant lobe, peaking at around 0.5$^\circ$ (30 kpc in projection) from
the active nucleus, that has no counterpart on the southern side.
High-resolution radio data (e.g. Junkes \etal\ 1993; Fig.
\ref{nml-fig}) show a high-surface-brightness feature, centred at
around RA = 13$^{\rm h}$ 27$^{\rm m}$ Dec = $-42^\circ$ $40'$, which
extends smoothly northwards into the larger-scale lobe, with the
magnetic field vectors oriented along the long axis of the structure
(Junkes \etal\ 1993). At higher resolutions still the brightest part
of the NML has been imaged by Morganti \etal\ (1999), who see a
jet-like structure (the `large-scale jet') connecting the NML and the
northern inner lobe. Although high-resolution imaging of the northern
inner lobe (e.g. Clarke \etal\ 1992) shows a sharply bounded radio
structure with no evidence for continued outflow, it is possible (as
argued by Morganti \etal) that the existence of the large-scale jet
implies that there is continued energy supply to the NML. This
interpretation would be consistent with the flat spectrum between 5
and 90 GHz of the base of the NML region (see above). It would also be
consistent with the detection of X-ray emission apparently embedded in
the lobes (Feigelson \etal\ 1981; Morganti \etal\ 1999) which is shown
by {\it XMM} observations to be highly overpressured thermal material,
plausibly requiring an interaction with an ongoing outflow (Kraft
\etal\ 2008, submitted to ApJ.). The NML is a highly complex region
associated with optical filaments and kinematically disturbed neutral
hydrogen (Morganti \etal\ 1999; Oosterloo \& Morganti 2005) and any
interpretation is difficult, particularly as we are not aware of
analogous structures in other restarting sources (Section
\ref{intro}). However, our data support the idea that the NML is, or
at least was until recently, fed by an active jet. Assuming an
equipartition magnetic field\footnote{Here and throughout the paper,
except where otherwise stated, we assume an electron-positron plasma
with no significant energetic contribution from protons.}, an
`injection index' (i.e. the synchrotron spectral index of the
low-energy electron population) of $\alpha = 0.5$ and a Jaffe \&
Perola (1973: hereafter J-P) ageing spectrum, the data require
particle acceleration to have ceased no more than $7 \times 10^6$
years ago.

\section{Spectra and spectral ageing in the giant lobes}
\label{giant}

As shown above, the spectra of the giant lobes of Cen~A present a
mixed picture. The southern giant lobe has a steep high-frequency spectrum
and (within the limitations of our data) the spectra appear to steepen
systematically both with frequency and with distance from the nucleus
(Figs \ref{spix}, \ref{spectrum}). In the northern giant lobe the spectra
seem very uniform as a function of position and do not steepen
significantly with frequency above 5 GHz. What causes these
differences in the two giant lobes?

\begin{figure}
\epsfxsize 8.5cm
\epsfbox{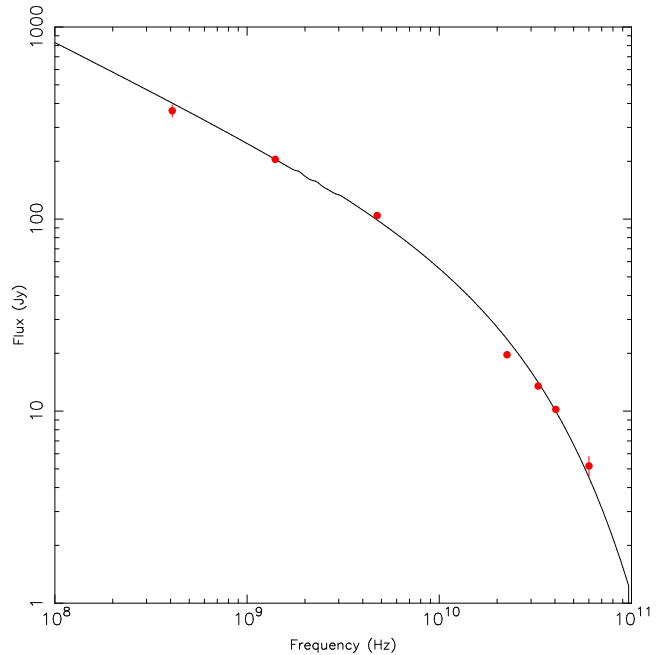}
\caption{Flux densities for region 4 (Table \ref{fluxes}) fitted with
  a standard Jaffe-Perola (1973) aged synchrotron spectrum assuming an
  injection index (low-frequency spectral index) of 0.5, an
  ageing $B$-field of 1.3 $\mu$G and a spectral age of $2.4 \times 10^7$ years.}
\label{age}
\end{figure}

We begin by noting that, given the large integration regions, the
southern giant lobe spectra are remarkably well fitted by standard J-P
aged synchrotron spectra with an injection index of $\alpha = 0.5$
(determined from the low-frequency spectral index)\footnote{We note
that the injection indices we assume are significantly flatter than
the spectral index of the inner lobes measured by Alvarez \etal . This
is not necessarily a problem, since 1) the equipartition magnetic
field strength in the inner lobes is much higher and 2) there is no
particular reason to believe that the particle acceleration process
currently operating in the inner lobes is the same as that which
operated in the giant lobes.}. We estimate an equipartition field
strength in the integration regions around 1.3 $\mu$G (treating them
as cylinders in the plane of the sky), which implies that
inverse-Compton losses dominate, since the CMB has an energy density
equivalent to a magnetic field with a strength of 3.3 $\mu$G. If we
assume that the field strength has the equipartition values and that
the CMB is the dominant photon field, the spectral fits (see Fig.\
\ref{age} for an example) imply spectral ages of $(2.4 \pm 0.1) \times
10^7$ years for region 4 and $(2.9 \pm 0.1) \times 10^7$ years for
region 5 (errors are $1\sigma$ statistical errors for one interesting
parameter only: systematic errors are dominated by the choice of
injection index and ageing field strength). By contrast, regions 1 and
2 (corresponding to the northern giant lobe) are rather poorly fitted
by such models, since their low-frequency spectra are too steep and
their high-frequency spectra too flat. They are better fitted with
standard `continuous injection' broken power-law models with an
injection index of 0.5 steepening to 1.0 at high frequencies (Heavens
\& Meisenheimer 1986). If a J-P model is fitted the derived ages are,
unsurprisingly, of the same order of magnitude but, implausibly,
region 2 appears older than region 1 (ages $(2.1 \pm 0.1) \times 10^7$
and $(2.9 \pm 0.1) \times 10^7$ years for regions 1 and 2
respectively, where we derive the age uncertainties by rescaling the
errors on the data points so as to make the reduced $\chi^2$ of the
fit equal to 1). The ages we derive are similar to the upper limits
quoted by Alvarez \etal\ (2000), who used somewhat lower magnetic
field strengths.

One possible explanation for the spectral differences between the two
giant lobes is that the northern lobe has undergone some particle
injection event that the southern lobe has not. A natural explanation
is that this is connected to the existence of the NML (Section
\ref{nml}): in this model the NML would be currently (or very
recently) connected to the energy supply and injecting high-energy
electrons into the base of the giant lobe. In this picture, the bright
radio structure extending from the NML into the northern giant lobe on
larger scales could be interpreted as continued outflow of recently
accelerated electrons, although this would have to extend to 100-kpc
scales to fully explain the peculiar spectra of lobe regions 1 and 2.
The strong differences in the polarization structures of the giant
lobes (Cooper \etal\ 1965; Junkes \etal\ 1993), and in particular the
fact that the northern giant lobe is strongly polarized with the
magnetic field direction aligned along the long axis of the NML,
support this picture, as do differences in the fine structure of the
radio emission in the northern and southern giant lobes (Feain \etal\,
in prep).

If this is the case, then the spectral ages derived for the southern
giant lobe (where there is no evidence of a connection to the energy
supply) allow us to suggest that the last acceleration of the
electrons now dominating the synchrotron radiation in this region took
place around $3 \times 10^7$ years ago (with the usual large
systematic uncertainties due to the assumption of equipartition). This
does not seem unreasonable given that various estimates of the
dynamical age of the inner lobes give a few $\times 10^6$ years. The
dynamical age of the giant lobes, if they have expanded at speeds of
the order of the sound speed in the hot external medium, with $kT \sim
0.35$ keV (Kraft \etal\ 2002), is expected to be almost an order of
magnitude higher, but this is consistent if we assume that high-energy
particles continued to be supplied to the southern giant lobe by a (now
dissipated) jet throughout its lifetime.

\section{Prospects for inverse-Compton detections}
\label{ic}
\begin{figure*}
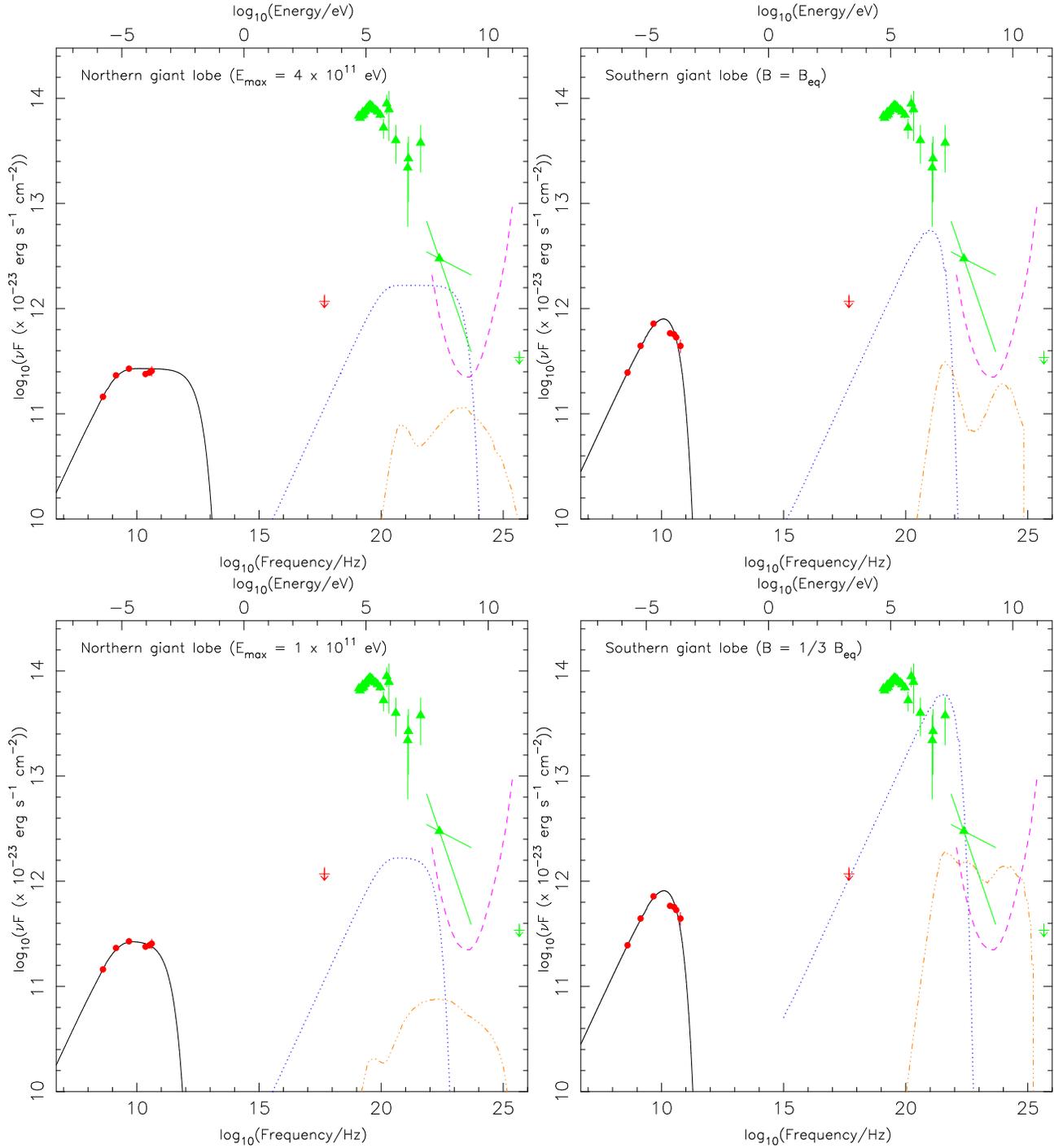

\epsfxsize 8.5cm
\epsfbox{cena-ngiantlobe.eps}
\epsfxsize 8.5cm
\epsfbox{cena-sgiantlobe.eps}
\epsfxsize 8.5cm
\epsfbox{cena-ngiantlobe-lc.eps}
\epsfxsize 8.5cm
\epsfbox{cena-sgiantlobe-mf.eps}
\caption{Inverse-Compton predictions for the giant lobes of Cen~A.
  Radio data points are the sum of our measurements from the
  ground-based radio maps and {\it WMAP} data for the regions
  corresponding to the northern (left) and southern (right) giant
  lobes as described in the text. High-energy data points are the same
  for all panels: the soft-X-ray limit of Marshall \&
  Clark (1981) is converted to a flux density at 2 keV ($\sim 5 \times
  10^{17}$ Hz) on the assumption that $\alpha = 0.5$. The 0.05 -- 30
  MeV ($\sim 10^{19}$ -- $5 \times 10^{22}$ Hz) points are the
  reported {\it Compton Gamma-Ray Observatory} ({\it CGRO}) OSSE and
  COMPTEL detections of Cen~A (Steinle et al. 1998), obtained from the
  NASA Extragalactic Database (NED), while the bow-tie plot centred on
  100 MeV is the EGRET spectrum reported by Sreekumar \etal\ (1999),
  and the point at 190 GeV ($5 \times 10^{25}$ Hz) represents the
  upper limit reported by Aharonian \etal\ (2005) from HESS
  observations, converted to a flux density by assuming power-law
  emission with photon index 3. We treat the high-energy data points
  as strict upper limits on the emission from the giant lobes. We also
  plot an estimate of the point-source {\it Fermi} sensitivity after
  one year (dashed line). The solid line shows the predicted
  synchrotron emission for the specified electron spectrum, the
  dotted line shows the inverse-Compton prediction for CMB photons,
  and the dot-dashed line shows the prediction for scattering of the
  EBL. For the northern giant lobe we plot the
  synchrotron and inverse-Compton spectra corresponding to a broken
  power-law electron energy spectrum with two possible high-energy
  cutoffs, $4 \times 10^{11}$ eV
  (upper panel) and $1 \times 10^{11}$ eV (lower panel), as discussed in the text. For the southern giant lobe
  (right-hand panel) we plot the synchrotron spectrum and
  inverse-Compton predictions for a J-P model as described in the text,
  with the upper panel showing the predictions for an equipartition magnetic field and the
  lower panel those for a field strength a factor 3 below equipartition.
}
\label{ic-predict}
\end{figure*}

The mixed results on the spectra of the giant lobes imply ambiguous
prospects for inverse-Compton detections at high energies. We estimate
the inverse-Compton spectrum for an equipartition magnetic field using
the code of Hardcastle, Birkinshaw \& Worrall (1998), taking the flux
densities of the giant lobes to be the sum of the fluxes for regions 1 and 2
(north lobe) and 4 and 5 (south lobe) in Table \ref{fluxes}, and
treating the lobes as cylinders in the plane of the sky. We then use
two reference models for the electron energy spectrum: a J-P spectrum
with a spectral age of $3 \times 10^7$ years, as found for regions 4
and 5, for the south lobe, and a broken power law, which provides a
better fit to regions 1 and 2, for the north lobe.

The energetically dominant photon population, and therefore the
population that dominates the inverse-Compton luminosity, is provided
by the CMB. At the large distances of the Cen A giant lobes the energy
density in photons provided by the host galaxy of Cen A is negligible,
and we do not include it in our calculations; if any inverse-Compton
emission arises from scattering of the galactic photons it will
preferentially be seen at small distances from the nucleus and will be
hard to resolve from the nuclear high-energy emission. However, as
Georganopoulos \etal\ (2008) have shown, the ubiquitous extragalactic
background light (EBL) provides a second photon population that must
be taken into account in high-energy inverse-Compton calculations. The
spectral energy distribution of the EBL is not well known (indeed,
Georganopoulos \etal\ propose inverse-Compton observations as an
additional constraint on its properties) and so we must adopt a model
to make a prediction. We base our estimates of EBL-related
inverse-Compton emission on the spectral energy
distribution\footnote{See also http://www.desy.de/$\sim$mraue/ebl/ .}
derived by Raue \& Mazin (2008): this, while not a complete model, is
intended to be consistent with all the existing direct and indirect
limits. The model we use is significantly lower in photon energy
density at all wavelengths than the upper range of the models adopted
by Georganopoulos \etal\ (2008) and taken from Mazin \& Raue (2007).
Our inverse-Compton code includes an approximation to the
Klein-Nishina correction to the effective cross-section of the
electron, since these effects start to become significant when the
highest-energy photons in the EBL model (which cuts off at $\lambda =
0.1$ $\mu$m) are scattered by the highest-energy electrons in the
lobes, and also account for the kinematical upper limit on the energy
of scattered photons (i.e. $h\nu < \gamma m_{\rm e}c^2$). Since these
calculations are only approximate, the results at the very highest
energies should be taken as indicative only.

The results of our calculations are
shown in Fig. \ref{ic-predict}. For the south giant lobe, where the J-P
spectrum is used, we see a sharp cutoff in the CMB inverse-Compton
prediction between about $10^{21}$ and $10^{22}$ Hz (4-40 MeV). This
is as expected, since in the J-P spectrum there are no electrons above
$\sim 3 \times 10^{11}$ eV (corresponding to $\gamma = 6 \times 10^5$)
and significant depletion of the spectrum below that, and the gain in
energy for the inverse-Compton process is of order $\gamma^2$. Thus we
do not predict a {\it Fermi} detection for equipartition field
strengths for the south giant lobe, either via CMB photons or via the
EBL (Fig. \ref{ic-predict}, upper right panel).

On the other hand, for the broken power-law spectrum used to model the
north giant lobe, the maximum energy of the electrons is
unconstrained. If we were arbitrarily to assume a sufficiently large
value ($\ga 10^{13}$ eV) we would obtain a spectrum from scattering of
the CMB that does not cut off even in the {\it Fermi} band ($10^{22}$
-- $10^{25}$ Hz). However, the predicted spectrum would be
inconsistent with the 30 MeV -- 2 GeV EGRET detection of Cen A
(Sreekumar et al. 1999; Hartman et al. 1999). The derived EGRET
position is coincident with the nucleus so we can treat this measured
spectrum, particularly at low energies, as a strict upper limit on the
spectrum of either of the giant lobes, since the angular resolution of
EGRET is $\sim 1^\circ$ or greater at these energies (see e.g. fig.\ 1
of Funk et al. 2008). If we adopt a lower value of $E_{\rm max}$,
$\sim 4 \times 10^{11}$ eV, we obtain an inverse-Compton spectrum that
does not violate the EGRET limit but still allows a detection of
scattered CMB photons\footnote{We use {\it Fermi} 1-year $5\sigma$
point-source sensitivities taken from
http://www-glast.slac.stanford.edu/software/IS/glast\_lat\_performance.htm:
these are appropriate fiducial values in the absence of observational
constraints on {\it Fermi}'s real performance for an extended source.
Below 100 MeV, the background is expected to be high and our
extrapolation is based on the sensitivity curve for the "inner galaxy"
presented in Funk et al. (2008).} of emission from this process at the
extreme soft end of the {\it Fermi} band. This $E_{\rm max}$ is the
one used in Fig.\ \ref{ic-predict} (upper left panel). Although still
lower values of $E_{\rm max}$ are possible, it cannot be reduced too
much before it starts to predict an (unobserved) high-frequency cutoff
in the synchrotron spectrum. However, values of $\sim 10^{11}$ eV are
permitted by the data, and this predicts no emission in the {\it
Fermi} band (Fig. \ref{ic-predict}, lower left panel). Again, neither
of these models predicts strong emission from inverse-Compton
scattering of the EBL in the {\it Fermi} band if it is close to the
level of our adopted model, although it would be detected in the most
sensitive part of the band if its level were close to the upper value
used by Georganopoulos \etal\ (2008).

The inverse-Compton calculation in Fig.\ \ref{ic-predict} assumes
equipartition, while in general we find in the lobes of FRIIs that the
observed inverse-Compton emission exceeds the equipartition prediction
by a factor of a few (e.g. Croston \etal\ 2005). Reducing the magnetic
field strength both increases the number density of electrons (since
the synchrotron emissivity is fixed) and increases their maximum
energy (since the observed cutoff frequency is fixed). However, the
existing observations, particularly the soft X-ray limit of Marshall
\& Clark (1981), constrain the magnetic field to be no more than a
factor $\sim 3$ below equipartition. If we reduce the magnetic field
by this factor, then the inverse-Compton prediction implies
detectability at very soft energies of the scattered CMB even for the
southern giant lobe (Fig. \ref{ic-predict}, lower right panel),
although we emphasize that {\it Fermi}-LAT observations are
challenging in this energy range due to the decreased effective area
and an increase in the background diffuse emission. In addition we
expect to see quite significant, spectrally distinct emission from
inverse-Compton scattering of the EBL. This prediction may already
marginally conflict with the {\it CGRO} data, and more extreme
electron-density excesses over the equipartition values are certainly
ruled out both by the Marshall \& Clark limit and the {\it CGRO}
spectrum, but we can conclude that even for the southern giant lobe
there is a region of parameter space with $B<B_{\rm eq}$ where a {\it
Fermi} detection is possible.

The results to be expected from {\it Fermi} therefore depend
sensitively on the magnetic field strength in the lobes and on the
unknown high-energy electron spectral cutoff in the north giant lobe.
Fig. \ref{ic-predict} illustrates some of the possibilities.
If the magnetic field strength is significantly below equipartition,
or if the high-energy cutoff is high enough in the north lobe, then we
expect a {\it Fermi} detection of scattered CMB photons for our
adopted sensitivities, although in all cases the {\it Fermi} emission
must be very soft to avoid violating the limits imposed by EGRET
between 30 MeV and 1 GeV. If {\it Fermi} does detect the giant lobes,
then we will measure the magnetic field strength, and the
sensitivities are such that we should be able to observe in the {\it
Fermi} spectrum the difference in the high-energy electron spectra of
the two giant lobes implied by the {\it WMAP} data: the resulting
constraints on the magnetic field strength will then allow a
measurement of, or limit on, the spectral energy density of the
background of photons at higher energies (e.g. from the EBL) in the
lobes. If extended inverse-Compton emission is not detected, then {\it
Fermi} will set field-dependent limits on the high-energy cutoff in
electron energies in the northern giant lobe and will improve
significantly on the existing limits on magnetic field strength, but
the problem of detecting inverse-Compton emission from Cen~A, and thus
{\it measuring} the magnetic field strength in the giant lobes, will
remain unsolved.

\section{UHECRs from the giant lobes and their possible radiative
  signatures}

\subsection{Introduction}

As mentioned in Section \ref{intro}, at least 4 of the 27 PAO UHECR
events in the energy range $E_{\rm p} \sim (0.6-0.9) \times
10^{20}$\,eV (Abraham \etal\ 2007, 2008) may be associated with Cen~A
(e.g. Moskalenko \etal\ 2008). Although evidence has been presented that
the UHECRs include some heavier nuclei as well as protons (e.g. Unger
\etal\ 2007), the level of uncertainty in the composition does not
affect our order of magnitude estimates if, as we assume hereafter,
all the events are identified with protons. In this section we argue
that the giant lobes of Cen~A may confine such extremely relativistic
protons, and may indeed be the sites for their acceleration. Our
spectral analysis of the giant lobes suggests that there is ongoing
continuous particle (electron) acceleration in the northern part of
the source, requiring, in turn, the presence of magnetic turbulence
which can energize different cosmic ray species via stochastic
particle-wave interactions. In addition, the parameters of the giant
radio lobes of Cen~A estimated in the previous sections allow us to
consider the problem of acceleration of UHECRs in Cen~A in a more
quantitative way than has previously been possible.

\subsection{Acceleration of UHECRs}

The Larmor radius of a cosmic ray with energy $E_{\rm p} \equiv
E_{20}\,10^{20}$ eV gyrating in a magnetic field with intensity $B
\equiv B_{-6}\,10^{-6}$ G is $r_{\rm L} =
E_{\rm p}/e B \sim 100\,E_{20} B_{-6}^{-1}$ kpc. Thus, the giant lobes
of Cen~A, with an estimated equipartition magnetic field $B_{-6} \sim
1$ and a spatial scale $R \sim 100$ kpc, satisfy the standard Hillas
(1984) criterion for a possible source of UHECRs, since $r_{\rm L}
\leq R$ for $E_{20} \leq 1$. The lobes of radio galaxies are highly
magnetized, rarified cavities inflated by jets in the surrounding
medium, and are expected to contain a predominantly tangled magnetic
field component with the maximum wavelength of the turbulent modes
possibly as large as the scale of the system. Since the giant lobes are
plausibly close to equipartition between magnetic field and particle
energy density, as indicated by our analysis in Sections \ref{giant}
and \ref{ic},
the velocities of such modes are expected to be very close
to the speed of light. These are excellent conditions for the
efficient stochastic acceleration of protons injected by the jets
into, and confined within, the lobes, by means of resonant Fermi-type
processes. The shortest possible timescale for such acceleration
corresponds to Bohm-type diffusion (where the particle mean free path
is comparable to the particle gyro-radius: we assume this timescale in
what follows), and is roughly\footnote{In general we can write
      $t_{\rm acc} = r_{\rm L}/\eta c$, where the factor $\eta$ accounts for less than ideal efficiency of
  particle acceleration; see, e.g., Aharonian \etal\ (2002). In the specific case
  of stochastic acceleration by magnetic turbulence (e.g. Stawarz \&
  Petrosian 2008) the acceleration timescale (in the Bohm limit and
  with a turbulence spectrum $W(k)$ going as $k^{-1}$) is given by
$t_{\rm acc} = (v_{\rm A}/c)^{-2} (U_0/U_{\rm turb}) (r_{\rm L} / c)$
where $v_{\rm A}$ is the Alfv\'en speed and $U_0/U_{\rm turb}$ is the ratio of
energy densities in the unperturbed magnetic field and the turbulent
magnetic component. Thus for relativistic strong turbulence ($v_{\rm
  A} \la c$, $U_{\rm turb} \la U_0$), we expect the factor $1/\eta \sim 10$.}
$t_{\rm acc} \sim 10 \, r_{\rm L} / c$. This implies that
\begin{equation}
t_{\rm acc} \sim 3.5\,E_{20}\,B_{-6}^{-1}\ {\rm Myr}
\end{equation}
which is comfortably
smaller than the age of the lobes estimated in the previous section,
$t_{\rm lobe} \sim 30$ Myr. On the other hand, the characteristic
timescale for the diffusive escape of particles from the lobes is of
the same order or longer only if $E_{20} \leq 1$, namely
\begin{equation}
t_{\rm diff} \sim 3\,R^2 / r_{\rm L} c \sim 0.9\,R_{100}^2\,E_{20}^{-1}\,B_{-6}
  \ {\rm Myr}
\label{eq:tdiff}
\end{equation}
where $R_{100} \equiv R/100$ kpc. The maximum energy of cosmic
rays available in the acceleration scenario we are considering, given
by the condition $t_{\rm acc} \sim t_{\rm diff}$ (since all the other
loss timescales are much longer than $t_{\rm diff}$, see below), is
$E_{20} \sim 0.5 \,B_{-6}\,R_{100}$ (thus satisfying the Hillas
criterion, since the condition $t_{\rm acc} \sim t_{\rm diff}$
reduces exactly to $r_{\rm L} \sim R$ for the assumed acceleration
and diffusion timescales). For the estimated lobe parameters $B_{-6}
\sim 1$ and $R_{100} \sim 1$, this maximum energy is in very good
agreement with the observed energies of the PAO events associated with
Cen~A (see above).

We next consider whether Cen~A is powerful enough to account for the
detected flux of UHECRs. We begin by estimating the energetics of the
giant lobes assuming rough equipartition between the magnetic field,
relativistic electrons and (not necessarily relativistic) protons. For
a cylindrical volume matching the giant radio structure with $V =
\pi\,R^2\,h \sim 5.5 \times 10^{71}\,R_{100}^2\,h_{600}$ cm$^3$ (the
radius $R \equiv R_{100}\,100$ kpc and height $h \equiv
h_{600}\,600$ kpc), $E_{\rm tot} \sim 3 \, U_{\rm B} \, V \sim 6.6
\times 10^{58}\,B_{-6}^2\,R_{100}^2\,h_{600}$ erg, where $U_{\rm B}
\equiv B^2/8 \pi \sim 4 \times 10^{-14}\,B_{-6}^2$ erg cm$^{-3}$ is
the magnetic energy density. This leads to a very rough limit to the
power of the jets supplying energy to the lobes
\begin{equation}
L_{\rm j} \sim
E_{\rm tot} / 2\,t_{\rm lobe} \sim 3.5 \times
10^{43}\,B_{-6}^2\,R_{100}^2\,h_{600}\ {\rm erg\,s}^{-1}
\end{equation}
This estimate should be considered as a lower limit, as it neglects any mechanical
work done by the jets on the surrounding medium and the actual jet
lifetime may also be substantially shorter than the age of the giant
lobes; however, it is of the same order of magnitude as kinetic power
estimates for the jets powering the inner lobes of Cen A (see Kraft
\etal\ 2003) and with jet power estimates for more distant FRI radio
galaxies (e.g. Laing \etal\ 2002). The UHECR flux detected
by PAO from a point source with $N$ associated events above $E_{\rm
th} = 60$ EeV energies, assuming a power-law form of the cosmic ray
spectrum with a energy index $s=2$, is $F_{\rm UHE} \sim 0.33
\times 10^{17}\,N\,E_{20}^{-2}$ km$^{-2}$ yr$^{-1}$ eV$^{-1}$, when
corrected for the relative exposure appropriate for Cen~A's
declination (e.g. Cuoco \& Hannestad 2008). The particle
energy index $s$ is poorly known in the energy range we are considering; for
simplicity, we assume hereafter the standard value of $s=2$,
consistent with the electron injection spectrum within the giant lobes
of Cen~A used in our analysis, which results in equal power stored
per decade of cosmic ray energy. For $N=4$, the monochromatic particle
energy flux is then
$[E^2_{\rm p}F]_{\rm UHE} \sim 0.66 \times
10^{-12}$ erg cm$^{-2}$ s$^{-1}$ and the monochromatic particle
luminosity is thus
\begin{equation}
L_{\rm UHE} \sim 4 \pi\,d^2\,[E^2_{\rm p}F]_{\rm UHE} \sim
10^{39}{\rm\ erg\,s}^{-1}
\label{eq:luhe}
\end{equation}
Even if we extrapolate the cosmic ray spectrum down
to the lowest energies with $s=2$, the total cosmic ray luminosity
is found to be $L_{\rm tot} \sim 10\,L_{\rm UHE} \sim
10^{40}$ erg s$^{-1}$, and thus it constitutes a negligible fraction ($\la
0.1\%$) of the power supplied by the jets to the giant lobes.

\subsection{Radiative signatures}

As discussed in the previous subsection, the timescales and energetics
are both consistent with the hypothesis that UHECRs are accelerated
continuously within the giant lobes of Cen~A. In this scenario, the
ultrarelativistic protons within the extended and magnetized lobes
will radiate, with the main loss processes being due to synchrotron
radiation and proton-proton ($p-p$) interactions\footnote{UHECR will
  also interact with the photon field within the lobes, via both 
  photo-meson production and Bethe-Heitler pair production. The key
  quantity for the former process is the number density of photons with energies 
above the photo-meson production energy threshold, $300$ MeV in the proton 
rest frame, which correspond to an observed photon energy $\varepsilon^{\star} 
= 3\,E_{20}^{-1}$\,meV, or a frequency $\nu^{\star} = 7 \times 10^{11} E_{20}^{-1}$
Hz (Aharonian 2002). However, for $E_{20} \sim 1$, the numerically
dominant photon field in the Cen A lobes around and above $\nu^{\star}$ (by
a very large factor) is the CMB, and it is well known (e.g.\ Greisen
1966) that the loss timescale is $>10^7$ years even for $E_{20} \sim
1$, and much larger for lower energies. Although the number density of
synchrotron photons available for pair production for a given UHECR energy is higher (because
of the lower energy threshold for pair production) CMB photons are
still numerically dominant and we know that pair production is
negligible compared to photo-meson production when CMB losses are
considered. We therefore neglect these loss processes
in what follows.}.  The former process
is characterized by a timescale
\begin{equation}
t_{\rm syn} \sim 1.4 \times
10^{6}\,E_{20}^{-1}\,B_{-6}^{-2}\ {\rm Myr}
\label{eq:tsyn}\end{equation}
resulting in a synchrotron
continuum peaking at photon energies $\varepsilon_{\rm syn} \sim
25\,E_{20}^2\,B_{-6}$ keV (Aharonian 2002). In the latter case, the
characteristic timescale is
\begin{equation}
t_{\rm pp} \sim 1.7 \times 10^6\,n_{-4}^{-1}\ {\rm Myr}
\end{equation}
where $n_{\rm th} \equiv n_{-4}\,10^{-4}$
cm$^{-3}$ is the number density of the cold gas within the giant
lobes, and $\gamma$-rays with energies $\varepsilon_{\gamma} \leq 10$
EeV for the maximum proton energy $E_{\rm p} \sim 10^{20}$ eV will be
generated (see e.g. Aharonian 2002).

It follows immediately from the
above results (eqs \ref{eq:tdiff}, \ref{eq:luhe}, \ref{eq:tsyn}) that the 
synchrotron emission of the cosmic-ray protons
produced within Cen A's giant lobes is negligible, since it is
expected to peak around photon energies $\sim 10$ keV with a
luminosity of the order of
\begin{equation}
L_{\rm syn} \sim L_{\rm UHE}\,t_{\rm
diff}/t_{\rm syn} \sim 10^{33}\,R_{100}^2\,B_{-6}^3\ {\rm erg\,s}^{-1}
\end{equation}
This is orders of magnitude below the expected X-ray luminosity of the
giant lobes' electrons due to inverse-Comptonization of the CMB photon
field for $B_{-6} \sim 1$ ($\sim 2 \times 10^{41}$ erg s$^{-1}$), and
large increases in the magnetic field strength over the equipartition
value would be required for the proton synchrotron emission to become
dominant.

The expected $\gamma$-ray emission from $p-p$ interactions might,
however, be somewhat more promising in terms of allowing an
independent detection of the high-energy protons. We note first that
the timescale for the diffusive escape of cosmic ray protons from the
system becomes longer than the timescale for $p-p$ interactions below
energies $E_{\rm p} \sim 100\,R_{100}^2\,B_{-6}\,n_{-4}$ TeV. Thus,
all of the power channeled to such particles in the acceleration
process during the lifetime of the source is released as
$\gamma$-ray emission, below photon energies $\varepsilon_{\gamma}
\sim 10$ TeV, with an expected photon index $\Gamma_{\gamma} = s$
(Kelner, Aharonian \& Bugayov 2006). Unfortunately, the number density
of thermal protons within the lobes of radio galaxies is not known.
The lack of observed internal depolarization in low-frequency radio
observations (Cioffi \& Jones 1980) has been used to set
temperature-independent limits of the order of $10^{-4}$ cm$^{-3}$ for
extended components of other radio sources (e.g. Eilek \etal\ 1984;
Spangler \& Sakurai 1985), but this method depends on assumptions
about the field geometry (Laing 1984) and in any case the required
multi-frequency radio polarization observations of Cen A are not
available to us. Isobe \etal\ (2001) claim an {\it ASCA} detection of
soft X-ray emission coincident with the outer part of the giant lobes,
with $kT = 0.62$ keV: assuming an abundance of 0.1 solar, we find that
their quoted soft X-ray (0.5--2.0 keV) flux corresponds to a density
$n_{\rm th}=1.6 \times 10^{-4}$ cm$^{-3}$ if it comes from a uniform
thermal plasma internal to the lobe. This is a strict upper limit on
the gas density since emission may also be coming from gas external to
the giant lobes (increasing the abundance reduces the derived
density). Similarly, the Marshall \& Clark (1981) 2--10 keV flux limit
acts as a strict upper limit on the density of hot ($kT = 2.5$ keV)
gas in the giant lobes, giving $n_{\rm th} \sim 1.3 \times 10^{-4}$
cm$^{-3}$ if all the emission comes from a uniform thermal plasma with
the same abundance. These limits on densities look comparable to the
expected parameters of the intergalactic medium at hundred-kpc
distances from the centre of the galaxy group hosting Cen~A (i.e.,
from the Cen~A host galaxy), and hence we assume $n_{-4} \sim 1$ for
illustration, although we note that gas with this density must be
relatively cold if it is not to dominate the energy density of the
giant lobes.

With this value of $n_{-4}$, and with the model
cosmic ray spectrum $s=2$ anticipated above and normalized to the PAO
flux, the expected monochromatic $\gamma$-ray luminosity from
p-p interactions emitted over the whole volume of the extended giant
lobes at photon energies $\leq 10$\,TeV is
\begin{equation}
L_{\gamma} \sim L_{\rm UHE} \, t_{\rm lobe} / t_{\rm pp} \sim 10^{34} \,
n_{-4} \, {\rm erg \, s^{-1}} \, ,
\end{equation}
corresponding to a monochromatic (e.g., $10$\,TeV) flux energy density 
of only $[\varepsilon S(\varepsilon)]_{\gamma} \sim L_{\gamma} / 
4\pi\,d^2 \sim 10^{-17}$\,erg\,cm$^{-2}$\,s$^{-1}$.
This can be compared with the upper limit set on the TeV emission from
the nucleus of Cen A by the existing 4.2-h HESS observation, $F(>{\rm
0.19\,TeV}) < 5.68 \times 10^{-12}$ ph cm$^{-2}$ s$^{-1}$ (Aharonian
\etal\ 2005; cf.\ Fig.\ \ref{ic-predict}), which, with a $\gamma$-ray
photon index $\Gamma_{\gamma} = 2$, corresponds to a monochromatic
luminosity of $[\varepsilon L(\varepsilon)]_{\rm 0.19\,TeV} < 2.8
\times 10^{39}$ erg s$^{-1}$, or a flux of $1.7 \times 10^{-12}$ erg
cm$^{-2}$\,s$^{-1}$. Thus it seems that the expected $\gamma$-ray 
luminosity is much below the sensitivity limit of any current TeV 
instrument. Note that in addition
the angular resolution of HESS in the $0.1-100$ TeV photon energy
range is $\la 0.1^\circ$, which is much better than the $\sim
10^\circ-1^\circ$ point spread function of EGRET (Section \ref{ic}) in
the photon energy range $\sim 0.03-1$ GeV. Cen A therefore does not
appear point-like to HESS, whereas it is effectively point-like at
least at low energies with EGRET, so that the sensitivity limit of
HESS appropriate for an extended source will be substantially higher
than the quoted point source value.

On the other hand, the estimated TeV flux of $\sim
10^{-17}$ erg cm$^{-2}$ s$^{-1}$ should be considered as a very
conservative lower limit. This is because we have assumed a rather
flat cosmic ray spectrum within the whole proton energy range. For any
more realistic (steeper) spectrum for the accelerated cosmic rays at
the highest energy range, the expected $\gamma$-ray flux increases.
For example, assuming $s=2.7$ (as used by other authors) above
some break energy $E_{\rm br} < E_{\rm th}$, and $s=2$ below $E_{\rm
br}$, we obtain the PAO flux at $E_{\rm p} > E_{\rm th}$ with $N=4$
events equal to
\begin{equation}
[E^2_{\rm p}F]_{\rm UHE} \sim 1.1 \times
10^{-12}\,(E_{\rm p}/{\rm 60\,EeV})^{-0.7}\ {\rm erg\,cm}^{-2}\,{\rm s}^{-1}
\end{equation}
(see Cuoco \& Hannestad 2008), and thus the monochromatic cosmic ray
luminosity $L_{\rm CR} \sim 1.8 \times 10^{39}\,(E_{\rm br}/{\rm
60\,EeV})^{-0.7}$ erg s$^{-1}$ at proton energies $E_{\rm p} < E_{\rm
br}$. For example, with $E_{\rm br} \sim 10^{18}$ eV (see in this
context Kachelriess \etal\ 2008), one obtains $L_{\rm CR} \sim 3.2
\times 10^{40}$ erg s$^{-1}$, which is more than one order of
magnitude higher than our previous estimate, resulting in an expected
monochromatic $\gamma$-ray energy flux at photon energies $\leq 10$\,TeV 
of the order of $[\varepsilon S(\varepsilon)]_{\gamma} \sim 3 \times 
10^{-16}$\,erg\,cm$^{-2}$\,s$^{-1}$. Yet another reason why this is a 
conservative lower limit is that the re-processing of higher-energy 
$\gamma$-rays, produced by the $E_{\rm p} > 100$ TeV protons has not 
been taken into account in our analysis, and this may increase the 
expected $\gamma$-ray flux by a factor of at least few. Still, in order
to reach the sensitivity limit of modern Cherenkov telescopes, the total
power channelled to the cosmic rays with energies $E_{\rm p} \leq 100$\,TeV
would have to be of the order of $10\%$ of the total jet power, $L_{\rm CR}
\sim 0.1 \, L_{\rm j}$, corresponding to a total energy stored in
cosmic rays from the lobes of $E_{\rm CR} \sim 10^{58}$\,erg.

The possibility of detecting the high-energy $\gamma$-ray emission
from $p-p$ interactions in the giant lobes of Cen A, even though not
particularly supported by our analysis presented above, is an exciting
one and worth some attention, since the large extent of these lobes
means that they can be resolved at the highest TeV $\gamma$-ray
energies using HESS. In this way the contribution from the inner
regions of Cen~A (including the inner, sub-pc to kpc scale jet, the
host galaxy, and active nucleus) to the total $\gamma$-ray signal
could be spatially resolved from the emission from the lobes. In this
context it is important to note that our analysis of the leptonic
emission generated within the extended lobes (Section \ref{ic})
indicates that we expect no significant TeV radiation produced by
ultrarelativistic electrons via inverse-Comptonization of either the
CMB or EBL photon fields, since the limits from synchrotron and
inverse-Compton continuum indicate a cut-off in the lobes' electron
spectrum around energies $E_{\rm e} \la 1$ TeV, providing an
effective upper limit on the energy of any inverse-Compton scattered
photons (Fig.\ \ref{ic-predict}). Thus, if any TeV emission is
detected from the giant lobes, it can be safely identified with being
hadronic in origin. This is in contrast to the situation at the
lower photon energy range probed by {\it Fermi}, which can also
resolve the giant lobes from the central regions of Cen~A, since the
expected leptonic (inverse-Compton) emission produced from the lobes
is expected to dominate at GeV photon energies over any hadronic
emission.

Obviously, UHECRs can in principle be accelerated in other components
of the very complex Cen~A system. Indeed, models for UHECR production
on the very smallest scales (around the SMBH or in the inner jet) have
been considered extensively in the literature. These models are all
based around the general idea that the accreting SMBHs in the centers
of at least some active galaxies provide a sufficient drop in
potential for such an extreme acceleration. Indeed, the
electromagnetic force associated with the black hole of mass $M_{\rm
BH}$ rotating with angular momentum $J$ in an external magnetic field
(supported by the accreting matter) with intensity $B_{\rm nuc}$, is
$\Delta V \sim J\,B_{\rm nuc}/M_{\rm BH}\,c$ (Phinney 1983). The
precise value for $B_{\rm nuc}$ is not known, but it is generally
expected that the nuclear magnetic field energy density cannot be
higher than (and may be equal to) the energy density of the accreting
matter (e.g. Ghosh \& Abramowicz 1997). Thus, for an object accreting
at some fraction $\eta$ of the Eddington rate
\begin{equation}
B_{\rm nuc} \sim 6
\times 10^4\,\eta^{1/2}\,M_{8}^{-1/2}\ {\rm G}
\end{equation}
where $M_8 \equiv M_{\rm BH}
/ 10^8\,M_{\odot}$. The spin of SMBHs in general is also a very poorly
known parameter. However, assuming the spin paradigm for efficient jet
production (Blandford 1990), one can expect SMBHs in elliptical-hosted
radio galaxies to be spinning at the maximal rate, $J \sim J_{\rm max}
\equiv M_{\rm BH}\,c\,r_{\rm g}$, where $r_{\rm G} = G\,M_{\rm
BH}/c^2$ is the Schwarzchild radius of the hole (see Sikora, Stawarz
\& Lasota 2007). These give $\Delta V \sim
10^{18}\,\eta^{1/2}\,M_8^{1/2}$ statvolt, and therefore the maximum
available energy for a test particle with charge $e$ accelerated in
this potential drop is as high as
\begin{equation}E_{\rm max} \sim 3 \times
  10^{20}\,\eta^{1/2}\,M_8^{1/2}\ {\rm eV}
\end{equation}
For $M_8 \sim 1$ in Cen~A (Marconi \etal\ 2006; H\"aring-Neumayer
\etal\ 2006) and assuming an accretion efficiency of order $\eta \sim
0.01$ (Evans \etal\ 2004), this leads to $E_{\rm max} \sim 3 \times
10^{19}$ eV, in agreement with the energies of the associated PAO
events. However, even in these models it is not clear where and how
the involved electric circuit closes. It may close very close to the
black hole, or along the edges of extended radio lobes (see e.g.
Blandford 2008). In the former case UHECR protons may indeed be
generated in the inner regions of Cen~A (see, in this context,
Kachelriess \etal\ 2008); however, in the latter case, stochastic
interactions with the magnetic turbulence taking place in the giant
lobes, as discussed in this paper (and by others, e.g. Fraschetti \&
Melia 2008), or some other (though more
speculative) mechanism involving reconnection of the lobes' magnetic
field (see in this context Benford \& Protheroe 2008), may mediate the
particle acceleration. Irrespective of the detailed mechanism, only in
the case of UHECR production in the giant lobes will there be a clear
radiative signature potentially detectable at GeV-TeV photon energies,
since any analogous emission produced in the inner regions of the
radio galaxy, even though possibly stronger due to the richer particle
and photon environment (and therefore more significant $p-p$ or
proton-photon interactions), will always be confused with other
$\gamma$-ray `nuclear' components (e.g., Chiaberge \etal\ 2001),
including direct emission from the inner jet, or the emission from a
giant pair halo created by the re-processing of such emission (as
discussed in the context of Cen~A by Stawarz \etal\ 2006). The
detection or non-detection of extended GeV-TeV emission from the giant
lobes of Cen A can therefore in principle help to distinguish between
these models.

\section{Summary and conclusions}

We have presented the first high-frequency, spatially resolved study
of the giant lobes of Centaurus A. We confirm recent findings (Israel
\etal\ 2008) that the overall lobe spectra steepen at frequencies
above 5 GHz, but in addition we have shown that the northern and
southern giant lobes are significantly different at these high
frequencies: the spectrum of the southern lobe steepens monotonically
(and is steeper further from the active nucleus) whereas the spectrum
of the northern lobe, after an initial steepening, remains consistent
with a power law to within the limits of our data. We suggest that
this indicates a real difference in the particle acceleration history
of the northern and southern giant lobes, perhaps due to the influence
of the poorly understood NML region.

Few FRI sources have so far been studied at radio frequencies above a
few tens of GHz, with more attention having been given to the hotspots
and jets of powerful FRIIs (e.g. Hardcastle \& Looney 2008 and
references therein). The {\it WMAP} data for Cen~A show that it is
both possible and useful to study the dynamics of FRIs with
high-frequency data. The Atacama Large Millimetre Array (ALMA), in
combination with current and next-generation telescopes operating at
lower frequencies, will allow spectral studies of large samples of
well-studied FRI radio galaxies to be carried out relatively easily.
It will be interesting to compare the results of such work with the
results presented here for Cen~A, and in particular to search for
other radio sources that show strong spectral differences between the
lobes at high frequencies.

Our results also have implications for the high-energy astrophysics of
Cen~A. If the simplest interpretation of our spectral modelling is
correct, then the southern giant lobe is a true relic with no ongoing
particle acceleration, while the northern giant lobe may have an
electron spectrum that extends to substantially higher energies. While
{\it Fermi} detections of both lobes remain possible, the northern
lobe is likely to be more clearly detectable, and all our {\it Fermi}
predictions imply a relatively soft spectrum for inverse-Compton
scattering of the CMB. Upper limits in the soft
X-ray and from the EGRET detector on board the {\it CGRO} already rule
out magnetic field strengths more than a factor of a few below
equipartition: {\it Fermi} will either detect the giant lobes or
substantially improve this limit.

Finally, we have shown above that the giant lobes of Cen~A are
potential sites for acceleration of protons to the highest observed
energies, $E_{\rm p} \sim 10^{20}$ eV, although we are required to
adopt the most optimistic acceleration timescale, corresponding to the
Bohm limit for the particle-MHD wave interactions, in order to be
efficient enough to account for the PAO UHECR flux. We have also
found that in this scenario the $\gamma$-ray radiative signatures
accompanying the acceleration process, resulting from the interactions
of ultrarelativistic protons with the thermal gas within the lobes,
can possibly be observed by Cherenkov telescopes and/or by {\it
Fermi}, thus permitting a test of this scenario.

\section*{Acknowledgements}

MJH thanks the Royal Society for a research fellowship. CCC is
supported by an appointment to the NASA Postdoctoral Program at
Goddard Space Flight Center, administered by Oak Ridge Associated
Universities through a contract with NASA. \L S acknowledges support
by MEiN grant 1-P03D-003-29. We thank Nils Odegard for providing the
{\it WMAP} data, Norbert Junkes for providing us with radio data from
ground-based observations of Cen~A, Matthieu Renard for helpful
discussions of the detectability of the Cen~A lobes with coded-mask
high-energy instruments, Gustavo Romero and Sergey Troitsky for
helpful comments on the initial version of the paper, Daniel Mazin and
Martin Raue for providing us with data on the EBL spectral energy
density, and an anonymous referee for valuable comments which have
allowed us to make significant improvements to the paper. This
research has made use of the NASA/IPAC Extragalactic Database (NED)
which is operated by the Jet Propulsion Laboratory, California
Institute of Technology, under contract with NASA.

\end{document}